\documentclass[a4paper,fleqn,usenatbib]{mnras}

\usepackage{txfonts} 

\usepackage[T1]{fontenc}
\usepackage{ae,aecompl}

\usepackage{graphicx}	
\usepackage{amssymb}	

\title[Discovery of Gamma Rays from 1ES\,1440+122]{Discovery of Very High Energy Gamma Rays from 1ES\,1440+122}

\author[S.~Archambault et al.]{
S.~Archambault$^{1}$,
A.~Archer$^{2}$,
A.~Barnacka$^{3}$,
B.~Behera$^{4}$,
M.~Beilicke$^{2}$,
W.~Benbow$^{5}$, \newauthor
K.~Berger$^{6}$,
R.~Bird$^{7}$,
M.~B\"{o}ttcher$^{31}$,
J.~H.~Buckley$^{2}$,
V.~Bugaev$^{2}$,
J.~V~Cardenzana$^{8}$, \newauthor
M.~Cerruti$^{5}$,
X.~Chen$^{9,4}$,
J.~L.~Christiansen$^{10}$,
L.~Ciupik$^{11}$,
E.~Collins-Hughes$^{7}$, \newauthor
M.~P.~Connolly$^{12}$,
W.~Cui$^{13}$,
H.~J.~Dickinson$^{8}$,
J.~Dumm$^{14}$,
J.~D.~Eisch$^{8}$,
M.~Errando$^{15}$, \newauthor
A.~Falcone$^{16}$,
S.~Federici$^{4,9}$,
Q.~Feng$^{13}$,
J.~P.~Finley$^{13}$,
H.~Fleischhack$^{4}$,
L.~Fortson$^{14}$, \newauthor
A.~Furniss$^{17}$,
G.~H.~Gillanders$^{12}$,
S.~Godambe$^{32}$,
S.~Griffin$^{1}$,
S.~T.~Griffiths$^{18}$,
J.~Grube$^{11}$, \newauthor
G.~Gyuk$^{11}$,
N.~H{\aa}kansson$^{9}$,
D.~Hanna$^{1}$,
J.~Holder$^{6}$,
G.~Hughes$^{4}$,
C.~A.~Johnson$^{17}$, \newauthor		
P.~Kaaret$^{18}$,
P.~Kar$^{19}$,
M.~Kertzman$^{20}$,
Y.~Khassen$^{7}$,
D.~Kieda$^{19}$,
H.~Krawczynski$^{2}$, \newauthor
S.~Kumar$^{6}$,
M.~J.~Lang$^{12}$,
A.~S~Madhavan$^{8}$,
G.~Maier$^{4}$,
S.~McArthur$^{21}$,
A.~McCann$^{22}$, \newauthor
K.~Meagher$^{23}$,
J.~Millis$^{24,24}$,
P.~Moriarty$^{25,12}$,
T.~Nelson$^{14}$,
D.~Nieto$^{26}$, \newauthor
A.~O'Faol\'{a}in de Bhr\'{o}ithe$^{4}$,
R.~A.~Ong$^{27}$,
A.~N.~Otte$^{23}$,
N.~Park$^{21}$,
J.~S.~Perkins$^{28}$, \newauthor
M.~Pohl$^{9,4}$,
A.~Popkow$^{27}$,
H.~Prokoph$^{4}$,
E.~Pueschel$^{7}$,
J.~Quinn$^{7}$,
K.~Ragan$^{1}$,
J.~Rajotte$^{1}$, \newauthor
L.~C.~Reyes$^{10}$,
P.~T.~Reynolds$^{29}$,
G.~T.~Richards$^{23}$,
E.~Roache$^{5}$,
G.~H.~Sembroski$^{13}$, \newauthor
K.~Shahinyan$^{14}$,
A.~W.~Smith$^{19}$,
D.~Staszak$^{1}$,
K~Sweeney$^{33}$,
I.~Telezhinsky$^{9,4}$, \newauthor
J.~V.~Tucci$^{13}$,
J.~Tyler$^{1}$,
A.~Varlotta$^{13}$,
V.~V.~Vassiliev$^{27}$,
S.~P.~Wakely$^{21}$,
R.~Welsing$^{4}$, \newauthor
A.~Wilhelm$^{9,4}$,
D.~A.~Williams$^{17}$,
B.~Zitzer$^{30}$
\\
$^{1}${Physics Department, McGill University, Montreal, QC H3A 2T8, Canada}\\
$^{2}${Department of Physics, Washington University, St. Louis, MO  63130, USA}\\
$^{3}${Harvard-Smithsonian Center for Astrophysics, 60 Garden Street, Cambridge, MA 02138, USA}\\
$^{4}${DESY, Platanenallee 6, 15738 Zeuthen, Germany}\\
$^{5}${Fred Lawrence Whipple Observatory, Harvard-Smithsonian Center for Astrophysics, Amado, AZ 85645, USA}\\
$^{6}${Department of Physics and Astronomy and the Bartol Research Institute, University of Delaware, Newark, DE 19716, USA}\\
$^{7}${School of Physics, University College Dublin, Belfield, Dublin 4, Ireland}\\
$^{8}${Department of Physics and Astronomy, Iowa State University, Ames, IA 50011, USA}\\
$^{9}${Institute of Physics and Astronomy, University of Potsdam, 14476 Potsdam-Golm, Germany}\\
$^{10}${Physics Department, California Polytechnic State University, San Luis Obispo, CA 94307, USA}\\
$^{11}${Astronomy Department, Adler Planetarium and Astronomy Museum, Chicago, IL 60605, USA}\\
$^{12}${School of Physics, National University of Ireland Galway, University Road, Galway, Ireland}\\
$^{13}${Department of Physics and Astronomy, Purdue University, West Lafayette, IN 47907, USA}\\
$^{14}${School of Physics and Astronomy, University of Minnesota, Minneapolis, MN 55455, USA}\\
$^{15}${Department of Physics and Astronomy, Barnard College, Columbia University, NY 10027, USA}\\
$^{16}${Department of Astronomy and Astrophysics, 525 Davey Lab, Pennsylvania State University, University Park, PA 16802, USA}\\
$^{17}${Santa Cruz Institute for Particle Physics and Department of Physics, University of California, Santa Cruz, CA 95064, USA}\\
$^{18}${Department of Physics and Astronomy, University of Iowa, Van Allen Hall, Iowa City, IA 52242, USA}\\
$^{19}${Department of Physics and Astronomy, University of Utah, Salt Lake City, UT 84112, USA}\\
$^{20}${Department of Physics and Astronomy, DePauw University, Greencastle, IN 46135-0037, USA}\\
$^{21}${Enrico Fermi Institute, University of Chicago, Chicago, IL 60637, USA}\\
$^{22}${Kavli Institute for Cosmological Physics, University of Chicago, Chicago, IL 60637, USA}\\
$^{23}${School of Physics and Center for Relativistic Astrophysics, Georgia Institute of Technology, 837 State Street NW, Atlanta, GA 30332-0430}\\
$^{24}${Department of Physics, Anderson University, 1100 East 5th Street, Anderson, IN 46012}\\
$^{25}${Department of Life and Physical Sciences, Galway-Mayo Institute of Technology, Dublin Road, Galway, Ireland}\\
$^{26}${Physics Department, Columbia University, New York, NY 10027, USA}\\
$^{27}${Department of Physics and Astronomy, University of California, Los Angeles, CA 90095, USA}\\
$^{28}${N.A.S.A./Goddard Space-Flight Center, Code 661, Greenbelt, MD 20771, USA}\\
$^{29}${Department of Applied Science, Cork Institute of Technology, Bishopstown, Cork, Ireland}\\
$^{30}${Argonne National Labouratory, 9700 S. Cass Avenue, Argonne, IL 60439, USA}\\
$^{31}${Centre for Space Research, North-West University, Potchefstroom, 2520, South Africa}\\
$^{32}${Astrophysical Sciences Division, Bhabha Atomic Research Centre, Trombay, Mumbai 400085, India}\\
$^{33}${Department of Physics and Astronomy, 251B Clippinger Research Laboratories, Ohio University, Athens, OH 45701, USA}\\
}

\date{Accepted XXX. Received YYY; in original form ZZZ}

\pubyear{2015}

\begin{document}
\label{firstpage}
\pagerange{\pageref{firstpage}--\pageref{lastpage}}
\maketitle


\begin{abstract}

The BL Lacertae object 1ES\,1440+122 was observed in the energy range from 85\,GeV to 30\,TeV by the VERITAS array of imaging atmospheric Cherenkov telescopes.  The observations, taken between 2008 May and 2010 June and totalling 53~hours, resulted in the
discovery of $\gamma$-ray emission from the blazar, which has a redshift $z$=0.163. 1ES\,1440+122 is detected at a statistical
significance of 5.5 standard deviations above the background with an integral 
flux of (2.8$\pm0.7_{\mathrm{stat}}\pm0.8_{\mathrm{sys}}$)~$\times$~10$^{-12}$~cm$^{-2}$ s$^{-1}$ (1.2\% of the
Crab Nebula's flux) above 200\,GeV. The measured spectrum is described well by a power law from
0.2\,TeV to 1.3\,TeV with a photon index of 3.1 $\pm$ 0.4$_{\mathrm{stat}}$ $\pm$ 0.2$_{\mathrm{sys}}$.
Quasi-simultaneous multi-wavelength data from the \textit{Fermi} Large Area Telescope (0.3--300~GeV) and the \textit{Swift} X-ray Telescope (0.2--10 keV) are additionally used to model the properties of the emission region.
A synchrotron self-Compton model produces a good representation of the multi-wavelength data.  Adding an external-Compton or a hadronic component also adequately describes the data.  

\end{abstract}




\section{Introduction}

Active galactic nuclei (AGN) are observed to emit electromagnetic radiation from radio waves up to very high energy (VHE; $E>100$~GeV) $\gamma$-rays. These objects, 
which make up only a small fraction of the total number of observed galaxies, 
are very luminous, extremely compact, and can exhibit large luminosity variability. 
Although AGN differ widely in their observed characteristics, a unified picture has emerged in 
which AGN are powered by accretion onto a super-massive black hole (10$^{7}$--10$^{9}~$M$_{\odot}$).
Near the black hole is a hot accretion disc surrounded by a thick torus of gas and dust.
In some AGN (the radio-loud population, $\sim$15\%), 
a highly-relativistic outflow of energetic particles form a highly collimated jet generating non-thermal emission.  
Blazars are thought to be the case where the jet is aligned with our line of sight~\citep{Urry:1995mg}.  

Blazar spectral energy distributions (SEDs) are dominated by non-thermal radiation.
This emission has a flat radio spectrum, radio and optical polarisation, and is often highly variable.  
1ES\,1440+122 belongs to the BL Lacertae (BL Lac) subclass of blazars.  
BL Lacs do not have broad emission lines present, unlike flat-spectrum radio quasars.  
Blazar SEDs are characterised by two broad peaks, with a significant fraction of the power often being emitted in the $\gamma$-ray band. The low-energy peak in the SED is well understood as synchrotron emission from relativistic electrons. 
However, there are competing models to explain the high-energy peak emission as dominated by  
either leptonic or hadronic interactions 
\citep{Blandford95, Bloom96, Mannheim98, Pohl00, 2000A&A...353..847A}. 
BL Lacs have been further classified depending on the position of their 
lower-energy peak.  \citet{Padovani:1994sh} originally proposed two classes.  Class definitions were extended to include an intermediate case by \citet{Nieppola06} and \citet{Abdo:2009iq}, though there is not agreement on where to place the boundaries between classes.
Based on a parabolic fit in log-log space to archival data, \citet{Nieppola06} determined the location of synchrotron peak in 1ES\,1440+122 to be at $\nu_{{\rm peak}}=10^{16.4}$\,Hz, lying near the border between their intermediate-frequency-peaked and high-frequency-peaked BL Lac (IBL and HBL, respectively) class definitions.  According to the classification scheme in \citet{Abdo:2009iq}, this synchrotron-peak frequency sets the classification of the source as a high synchrotron peaked (HSP) BL Lac.  

1ES\,1440+122 was initially classified as an AGN in X-rays in the Einstein Slew Survey \citep{Elvis92}. It is surrounded by $\sim$20 galaxies within approximately 200\,kpc~\citep{Heidt99}, which suggests that it may belong to a small cluster of galaxies. Indeed it is likely that the blazar is interacting with an elliptical galaxy with a projected separation of $\sim$4\,kpc \citep{Sbarufatti06}. The host galaxy has been resolved in several imaging studies, and high-resolution HST imaging \citep{Scarpa99} reveals a very close companion ($\sim0.3^{\prime\prime}$) now known to be a foreground star \citep{2004ApJ...613..747G}. The optical spectrum of 1ES\,1440+122 is well measured, and a redshift of $z$=0.163 is obtained from the identification of three spectral lines \citep{Sbarufatti06}.

1ES\,1440+122 was identified as a likely VHE emitter on the basis of its SED \citep{Costamante02}.
It was observed by the H.E.S.S. array of imaging atmospheric Cherenkov telescopes (IACTs) for 11.2\,hr between 2004 and 2009 resulting in a 99.9\% confidence level integral flux upper limit of 1.66~$\times$~10$^{-12}$~cm$^{-2}$ s$^{-1}$ (1.0\% of the Crab Nebula's flux) above 290\,GeV~\citep{2014A&A...564A...9H}. 1ES\,1440+122 belongs to the 3FGL catalogue 
(3FGL J1442.8+1200) with spectral index, $\Gamma$ = 1.80 $\pm$ 0.12 \citep{2015ApJS..218...23A}. 
The \textit{Fermi} Large Area Telescope (LAT) spectrum shows no sign for a cutoff, and its extrapolation to the VHE band predicts a $\sim$2\% Crab Nebula flux between 200\,GeV and 1\,TeV, including extragalactic background light (EBL) absorption effects \citep{Franceschini08}. 
The 3FGL lists the source as being non-variable \citep{2015ApJS..218...23A}.  

VHE $\gamma$-ray emission from 1ES\,1440+122 was discovered by VERITAS~\citep{Ong10} during the 2008--2010
observing seasons. After the confirmation of the initial excess from this source, 
\textit{Swift} X-ray Telescope (XRT) target of opportunity observations were triggered in order to provide a detailed X-ray spectrum, which is crucial for constraining models of emission.  
In combination with \textit{Fermi}-LAT data, all data were used to construct an SED spanning 10 decades in energy.

\section{VERITAS observations}

VERITAS consists of four 12-m diameter IACTs located in southern Arizona \citep{Holder:2006gi}.  The array detects $\gamma$-ray emission from astrophysical objects in the energy range from $\sim$85\,GeV to $\sim$30\,TeV. 
VERITAS has an energy resolution of $\sim$15\% and angular resolution (68\% containment) 
of $\sim$0.1$^{\circ}$ per event. The current sensitivity of the array allows for a 1\% Crab Nebula flux source detection
in 25\,hours (5$\sigma$ detection), while a 10\% Crab Nebula flux is detected 
in 0.5\,hours.  Note that for sources with a softer spectrum than the Crab, the observing time required for detection will be longer.  The field of view of the VERITAS telescope has a diameter of 3.5$^{\circ}$. 
More information on VERITAS and the IACT technique can be found in~\citet{Holder08}. 

VERITAS observed 1ES\,1440+122 for about 78\,hours from 2008 May to 2010 June, 
which includes two observing seasons, 2008--2009 and 2009--2010. The earlier data were taken with the original VERITAS telescope configuration,
while the later data were taken after one of the telescopes was relocated in order to increase the array sensitivity \citep{Perkins:2009ia}. These observations cover the zenith angle range from 19$^{\circ}$ to 38$^{\circ}$. 
All the observations were performed in a mode where the source is offset by 0.5$^{\circ}$ from the centre of the field of view.  This offset allows for simultaneous background estimation with good precision while maintaining a high signal efficiency.  This is known as ``wobble'' mode \citep{Fomin1994}.  
Observations affected by poor weather or hardware problems were removed, and the remaining 53\,hours of data were processed 
with two independent analysis packages \citep{Daniel07} yielding consistent results. 

Images of the showers were first calibrated in gain and timing at the pixel level using nightly calibration data using an artificial light source. 
Following calibration, images from each telescope were parameterized using fits to two-dimensional Gaussian distributions.  
This technique is similar to the frequently-used moment analysis \citep{Hillas85}. 
Tests performed using $\gamma$-ray simulations have shown that the use of the Gaussian fit leads to several improvements with respect to the moment analysis.  
Truncated images, where some part of the shower is not contained in the field of view of the camera, are reconstructed with better angular and energy resolution. 
This also leads to an increase in the rate of events passing selection at high energy.  
No image cleaning (thresholding) was used, but a constant offset represents the night-sky background.  
As a result, background rejection was improved at low energy.  
More details of this image-fitting technique are given in \citet{HFit12}.

From the results of the image fits, parameters were calculated and used for event reconstruction and selection. The event selection criteria were optimised beforehand using observations of the Crab Nebula with the excess counts scaled to match a 1\% Crab Nebula flux source.  The source region was defined by a 0.1$^{\circ}$ radius circle centred on the source coordinates, and all the $\gamma$-ray-like events within this region were considered the ON counts. The reflected-region model \citep{2007A&A...466.1219B} was used for background 
subtraction, where the background was estimated from eleven identically-sized regions 
reflected from the source region around the camera centre. The events found in these
regions were considered the OFF counts. The Li \& Ma 
Formula 17 \citep{LiMa83} was used to calculate the significance at the source location. 

\begin{figure}
\includegraphics[width=\columnwidth]{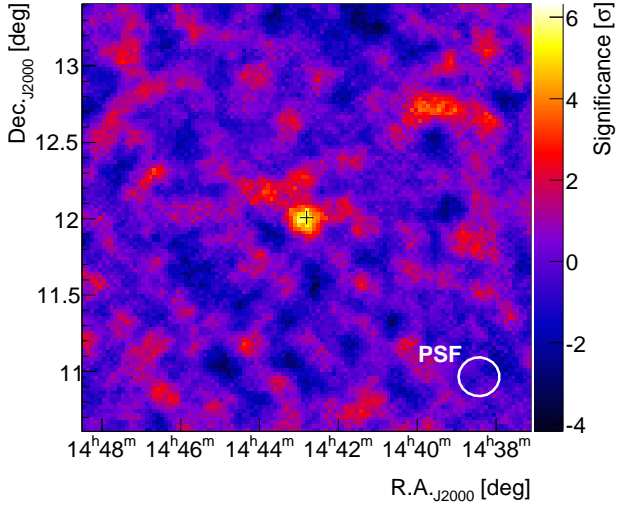}
\caption{VHE significance map of the region around 1ES\,1440+122 from the VERITAS observations.  The map has been smoothed using events within a radius of 0.1$^\circ$.  The black cross marks the location of 1ES\,1440+122 as reported in the SDSS.  The VERITAS angular resolution is indicated by the white circle.}
\label{skymap}
\end{figure}

\begin{figure}
\includegraphics[width=\columnwidth]{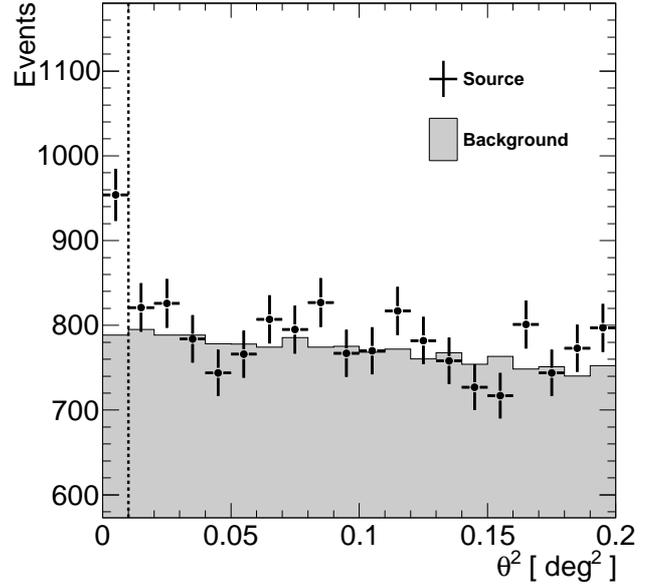}
\caption{Distribution of $\theta^{2}$ for the source (cross) and background regions (shaded region; normalised) from the VHE observations of 1ES\,1440+122.  $\theta$ is the angular distance between the source and the reconstructed event location.  The vertical dashed line indicates the ON region.}
\label{skymap2}
\end{figure}

An excess of 166 events was observed in VERITAS data from the direction of 1ES\,1440+122 
(954 ON events, 8673 OFF events with an off-source normalisation ratio of 
$\alpha = 0.0909$). The excess corresponds to a statistical significance of 5.5$\sigma$.
The significance map in the vicinity of 1ES\,1440+122 is shown in Figure~\ref{skymap} while the distribution of events with respect to the source location is shown in Figure~\ref{skymap2}.  
The VERITAS point spread function is 6$^{\prime}$ for 68\% containment radius at these zenith angles making the distribution of events consistent with a point-like source. 
Fitting a symmetric 2-dimensional Gaussian to the uncorrelated excess counts map results in a best-fit centroid at R.A. =
14$^{\rm{h}}$ 43$^{\rm{m}}$ 15$^{\rm{s}}$ and Dec. = +12$^{\circ}$ 00$^{\prime}$ 11$^{\prime\prime}$.  The new TeV source is catalogued as VER\,J1443+120.  The statistical uncertainty in the position of 1$^{\prime}$ and systematic uncertainty of 25$^{\prime\prime}$ make the $\gamma$-ray emission consistent with the position of 1ES\,1440+122, which is reported in the SDSS\footnote{\tt{http://www.sdss.org/dr5/}} as R.A. = 14$^{\rm{h}}$ 42$^{\rm{m}}$ 48.3$^{\rm{s}}$ and Dec. = +12$^{\circ}$ 00$^{\prime}$ 40$^{\prime\prime}$.
We fitted the integrated flux light curve $>$200~GeV in monthly bins with a constant-flux hypothesis, resulting in no statistically significant evidence of variability ($\chi^2$/dof =15.2/9 or a 9\% chance of being generated given a constant flux hypothesis).  

The spectrum of 1ES\,1440+122 is described well by a power law of the form $dN/dE = I_{0} \, (E/0.5~\mathrm{TeV})^{-\Gamma}$, with $I_{0} = (1.47 \pm 0.62_{\mathrm{stat}}) \times 10^{-12}$ cm$^{-2}$ s$^{-1}$\,TeV$^{-1}$ and $\Gamma = 3.1 \pm 0.4_{\mathrm{stat}}$, resulting in $\chi^{2}$/dof = 2.4/2.  

\begin{figure}
\includegraphics[width=\columnwidth]{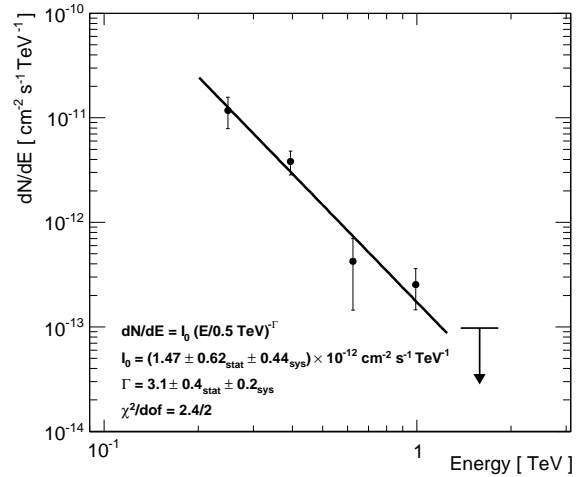}
\caption{VHE spectrum of 1ES\,1440+122 derived from the whole VERITAS data set.
An upper limit with 99\% confidence level is shown above the last significant spectral point.  }
\label{vhespectrum}
\end{figure}

The integral photon flux above 200\,GeV is $\Phi_{E > 200~\mathrm{GeV}} = (2.8 \pm 0.7) \times 10^{-12}$ cm$^{-2}$ s$^{-1}$.
Using the parameterization by \citet{2006A&A...457..899A}, this is equivalent to 1.2\% of the flux from the Crab Nebula above 200\,GeV.  
We estimate the systematic errors on the flux normalisation constant and the photon index to be $\Delta I_{0}$/I$_{0}$ = 30\% and $\Delta\Gamma = 0.2$. The VERITAS spectral points are shown in Figure~\ref{vhespectrum}.  
For a strict comparison with the H.E.S.S integral flux upper limit of 1.66~$\times$~10$^{-12}$~cm$^{-2}$ s$^{-1}$ above 290\,GeV~\citep{2014A&A...564A...9H}, we recalculate our integral flux above the same energy threshold to be $(1.5\pm 0.7)$~$\times$~10$^{-12}$~cm$^{-2}$ s$^{-1}$, which shows they are consistent with one another.

\section{\textit{Fermi}-LAT Observations}
The \textit{Fermi}-LAT is a pair-conversion $\gamma$-ray detector 
sensitive to photons in the energy range from about 30\,MeV to 300\,GeV \citep{2009ApJ...697.1071A}. 
The \textit{Fermi}-LAT data for this analysis were taken between 2008 August 04 and 2010 August 02. 
We used the likelihood tools distributed with the standard Science Tools v9r32p5 package available from the \textit{Fermi} Science Support Center\footnote{\tt 
http://fermi.gsfc.nasa.gov/ssc}. 
Events were required to have zenith angle $<$105$^{\circ}$ in order to limit contamination by the Earth albedo effect.  
Only events in the energy range of 300\,MeV to 100\,GeV and within a circular region 
of 12$^{\circ}$ radius centred on the source were selected.
The background was modelled with a galactic diffuse emission model\footnote{\tt gll\_iem\_v02.fit} and an isotropic 
component. Catalogued sources within 17$^{\circ}$ of the target location were included in 
the model. The fluxes were determined using the instrument response functions P7REP\_SOURCE\_V15 (a check was performed using updated response functions released after modeling was complete but these did not yield significant changes).  
The systematic uncertainty on the flux is approximately 5\% at 560\,MeV and under 10\% at 10\,GeV and above \citep{2012ApJS..203....4A}.

A point source was detected at the position of 1ES\,1440+122 with a significance of more 
than 9 standard deviations (TS$=$89.6). The highest-energy photon detected by the \textit{Fermi}-LAT with a high probability of association with the source (99.3\%) has an energy of 62\,GeV. The time-averaged \textit{Fermi}-LAT spectrum computed assuming a constant power law includes five points and is shown in Figure~\ref{fig_3models}.  The spectrum is fit by a power law that can be described as $dN/dE=N_{0}(1-\Gamma)E^{-\Gamma}/(E_{\mathrm{max}}^{1-\Gamma}-E_{\mathrm{min}}^{1-\Gamma})$, where $\Gamma=1.52\pm0.16_{\mathrm{stat}}$, $N_{0}=(5.39\pm1.18_{\mathrm{stat}})\times10^{-10}$ $\textrm{photons}$~$\textrm{cm}^{-2}$~$\textrm{s}^{-1}$, $E_{\mathrm{min}}=$1\,GeV and $E_{\mathrm{max}}=$100\,GeV.  
The values reported in the 3FGL are: $\Gamma=1.80\pm0.12$ and $N_{0}=(5.59\pm0.86)\times10^{-10}$ $\textrm{photons}$~$\textrm{cm}^{-2}$~$\textrm{s}^{-1}$ with the same min and max energy \citep{2015ApJS..218...23A}.  There is some tension between the spectral index we found and the 3FGL value.  The period of our observations matches the 2FGL catalogue more closely, and our spectral index is in agreement with the 2FGL value of $\Gamma=1.41\pm0.18$ \citep{Fermi-LAT:2011iqa}. 

\section{\textit{Swift}-XRT Observations}

\textit{Swift}-XRT \citep{Gehrels04} observations of 1ES\,1440+122 were  performed on 2008 June 12 and 2010 March 9.  
The \textit{Swift}-XRT data were analysed with HEAsoft\footnote{\tt http://heasarc.nasa.gov/lheasoft} 6.13. Both observations were completed in photon-counting mode, showing count rates of 0.40 $\pm$ 0.01 and 0.38 $\pm$0.02 counts s$^{-1}$. With these low count rates, 
photon pile-up is negligible and systematic uncertainties on the flux are negligible compared to the statistical errors.  

\texttt{XSPEC} version 12.8.0 was used for the XRT spectral analysis.  The data were combined and grouped into bins with a minimum of $20$ counts per bin, enabling the use of $\chi^2$ spectral-model fitting.   The time-averaged 0.3-10 keV data were fitted with an absorbed power-law model (\textit{tbabs(po)} in \texttt{XSPEC}) where the hydrogen column density $N_{H}$ was fixed at 1.58 $\times$ 10$^{20}$ cm$^{-2}$,
taken from the LAB survey of Galactic HI~\citep{Kalberla:2005ts}.
The data were reasonably well fit ($\chi^2/$dof $=61.9/45$) by an absorbed power law with normalisation at 1\,keV of  (3.2 $\pm$ 0.1)  $\times$ 10$^{-3}$\,s$^{-1}$\,keV$^{-1}$ and photon index of 1.95 $\pm$ 0.04.  In order to represent the intrinsic X-ray emission, the absorption-corrected spectrum was used for SED modelling.  The fluxes during both periods are consistent with one another, so an average was used.

\section{\textit{Swift}-UVOT Observations}

The Ultraviolet/Optical Telescope (UVOT; \citet{2005SSRv..120...95R}) onboard \textit{Swift} observed 1ES\,1440+122 simultaneously with the \textit{Swift}-XRT time periods.  Source photons in each of six filters (V, B, U, UVW2, UVM2, and UVW1) were extracted from a circular aperture of radius 5.0$^{\prime\prime}$ centred on the source.  The background was estimated in a 30$^{\prime\prime}$ radius circular region located away from the blazar.  The fluxes were computed using the \textit{uvotsource}\footnote{\tt{HEASOFT v6.13, Swift\_Rel4.0(Bld29)\_14Dec2012 with calibrations from \citet{2011AIPC.1358..373B}.}} tool.  Corrections for interstellar absorption were made using the extinction curve of \citet{1999PASP..111...63F} and assuming an $E(B-V)$ value of 0.0239$\pm$0.0006, determined from the IPAC Extragalactic Database~\citep{2011ApJ...737..103S}.  The average fluxes from the two observations were used for the SED modelling.

\section{Modelling and Discussion}

\begin{figure}
\includegraphics[width=\columnwidth]{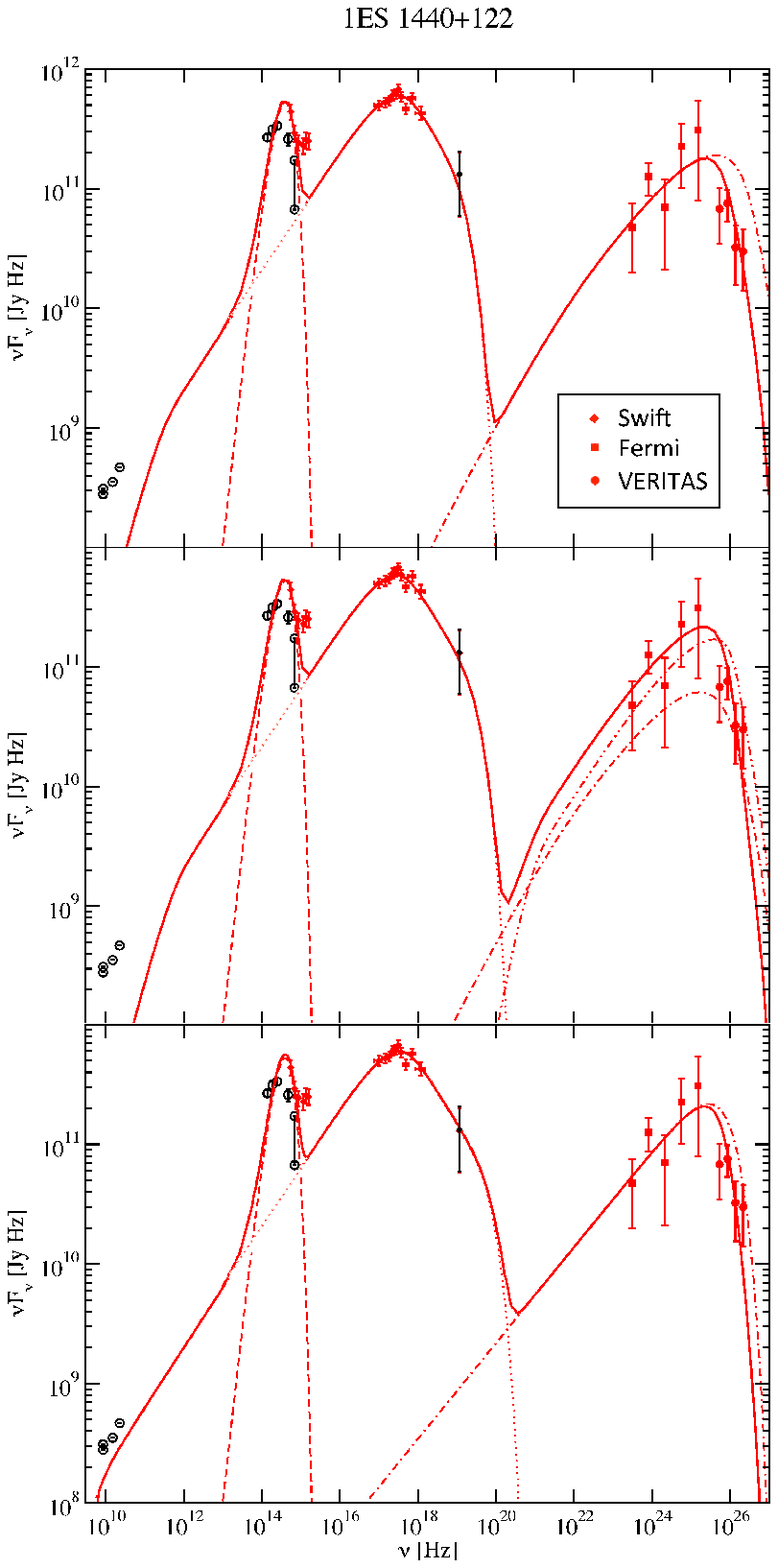}
\caption{\label{fig_3models}
SED of 1ES\,1440+122 using quasi-simultaneous \emph{Swift}, \emph{Fermi}-LAT and VERITAS data (red points), archival data (black points), and models (lines). 
The synchrotron component (dotted) and the total fluxes after correction for EBL absorption \citep{Franceschini08} are shown for every case (solid).  
The host galaxy was modelled as a blackbody (dashed).  
Top panel:  SSC model with inverse-Compton component (dot-dashed) shown.  
Middle panel: EC model with self-Compton  (dot-dashed) and inverse-Compton components (dot-dot-dashed) shown.
Lower panel: Hadronic model with hadronic component (dot-dashed) shown.  
}
\label{spectrum}
\end{figure}

The quasi-simultaneous SED of 1ES\,1440+122 
with data from VERITAS, \textit{Fermi}-LAT, and \textit{Swift}-XRT is shown in Figure 
\ref{fig_3models} and parameters of the three models, discussed below, are listed in Table~\ref{fitpars}. To these data, we added archival low-frequency radio and optical data from the NED\footnote{\tt http://nedwww.ipac.caltech.edu} as well as a \textit{Swift}-BAT point\footnote{\tt http://tools.asdc.asi.it}. The vertical bar between the two
B-band points illustrates the amount of historical variability at
that frequency. The IR-through-optical emission of 1ES\,1440+122 is
clearly dominated by thermal emission from the host galaxy, modelled as a blackbody spectrum here.  
The UV data are under-represented in all models that we investigated.  
This could be additional contamination from the host galaxy (whose emission has been parametrized by a simple blackbody though it is known to be more complex) or extended jets unrelated to the VHE emission.  The latter scenario has been considered for PKS\,2155-304, which does not show correlated variability between the two bands~\citep{Abramowski:2012ux}.  Optical polarisation data may be useful for disentangling contamination from other parts of the jet~\citep{deAlmeida:2014iua}.  

We produce models of the SED of 1ES\,1440+122 with both leptonic and hadronic jet models.  The VHE emission is corrected for EBL absorption according to \citet{Franceschini08}, which is in agreement with the most recent constraints from gamma-ray observations \citep{2013A&A...550A...4H, 2012Sci...338.1190A, 2015ApJ...812...60B}. 
In leptonic models for blazar emission, a population of relativistic electrons is responsible for both the lower-frequency component of the SED (via synchrotron emission) as well as the higher-frequency emission (via Compton scattering).  
Potential soft photon fields that can serve as targets for Compton
scattering are either the synchrotron photons
(SSC = synchrotron self-Compton), or radiation fields produced externally 
to the jet (EC = External Compton).  
For our models, we use a steady-state
scenario in the fast-cooling regime based on the time-dependent blazar jet radiation transfer code
of \citet{bc02}, as described in detail in \citet{2013ApJ...768...54B}. In this
model, the emission originates from a spherical region of
radius $R$, moving along the jet with a Lorentz factor $\Gamma$,
corresponding to a jet speed $\beta_{\Gamma} c$. The jet is oriented 
at an angle $\theta_{\rm obs}$ with respect to the line of sight, 
resulting in Doppler boosting characterised by the Doppler factor 
$D = (\Gamma [ 1 - \beta_{\Gamma} \cos\theta_{\rm obs}] )^{-1}$.

Non-thermal electrons are injected and accelerated into a power-law
distribution at a rate $Q(\gamma) = Q_0 \gamma^{-q}$ between a
low- and high-energy cutoff, $\gamma_{1,2}$.  
A value of $q=3.0$ was chosen for all models, though it is not well constrained by the observations and can take a wide range of values depending on the obliquity and shock velocity~\citep{Summerlin:2011xs}.  
An equilibrium between this particle injection, radiative cooling and particle escape is established self-consistently with the radiation mechanisms considered. 
Particle escape was parameterized through an escape parameter $\eta$ such that the escape time scale $t_{\rm esc} = \eta \, R/c$. 
The resulting particle distribution will correspond
to a power $L_e$ in electrons streaming along the jet 
\cite[see][]{Acciari:2009xz}. The synchrotron emission is evaluated
assuming the presence of a tangled magnetic field $B$, corresponding
to a power in Poynting flux, $L_B$. For each model calculation, our 
code evaluates the equipartition parameter $\epsilon_{Be} \equiv
L_B/L_e$. Because of a lack of observational constraints and in
order to reduce the number of free parameters, we choose the
observing angle as the critical angle, for which $\Gamma = D$,
i.e., $\cos\theta_{\rm obs} = \beta_{\Gamma}$. 

\begin{table}
\centering
\begin{tabular}{|c|c|c|c|}
\hline
Parameter & SSC & EC & lepto-had. \\
\hline
$L_e$ [erg s$^{-1}$] & $2.2 \times 10^{43}$ & $5.3 \times 10^{42}$ &   $1.4 \times 10^{40}$ \\
$\gamma_1$   & $1.5 \times 10^5$ & $1.0 \times 10^5$ & $1.4 \times 10^4$ \\
$\gamma_2$   & $1.0 \times 10^6$ & $1.0 \times 10^6$ & $2.0 \times 10^5$ \\
$q$          & $3.0$  & $3.0$             & $3.0$  \\
$B$ [G]      & $0.15$  & $0.5$             & $30$   \\
$\Gamma = D$ & $25$   & $15$              & $15$   \\
$\eta$ & 1000 & 100 & 3 \\
$R$ [cm] & $3.5 \times 10^{15}$ & $5 \times 10^{15}$ & $5 \times 10^{15}$ \\ 
$u_{\rm ext}$ [erg cm$^{-3}$]  & --- & $4 \times 10^{-6}$ & --- \\
$T_{\rm ext}$ [K]              & --- & $10^3$ & --- \\
$L_p$ [erg s$^{-1}$]           & --- & --- & $8.1 \times 10^{44}$ \\
$\gamma_{\rm p}^{\rm min}$     & --- & --- & $1.1 \times 10^3$               \\
$\gamma_{\rm p}^{\rm max}$     & --- & --- & $1.3 \times 10^{10}$            \\
$p$                            & --- & --- & $2.2$                \\
\hline
$\epsilon_{Be}$ & $2.9 \times 10^{-2}$ & $1.0$ & $1.4 \times 10^6$ \\
$\epsilon_{Bp}$ & ---                  & ---    & $24$ \\
$\epsilon_{ep}$ & ---                  & --- & $1.7 \times 10^{-5}$ \\
$t_{\rm var}^{\rm min}$ [hr]  & $1.7$ & $4.2$ & ---            \\
\hline
\end{tabular}
\caption{SED model parameters described in the text.\label{fitpars}}
\end{table}

In a pure SSC model, only synchrotron photons play the role as targets
for Compton scattering. 
The SSC model satisfactorily produces the
non-thermal SED with plausible parameters. The size of the emission
region used in the model implies a minimum variability time scale
allowed by the model given by $t_{\rm var}^{\rm min} = R(1+z)/(c D)
= 1.7$\,hours. The required magnetic field energy density is a factor of approximately 35 below equipartition with the non-thermal electron
distribution.

For a model including an external radiation field as target for
Compton scattering, we have improved the model presented in \citet{Acciari:2009xz} 
by allowing for isotropic (in the rest frame
of the AGN) radiation fields with arbitrary spectra. Guided by
recent results of EC modelling of SEDs of other VERITAS-detected
IBLs, such as W Comae \citep{Acciari:2008rk,Acciari:2009xz} and 3C66A
\citep{abdo11}, we consider a thermal infrared radiation field,
possibly originating in a dusty torus around the central engine,
as an appropriate choice for an external radiation field.  
The energy density $u_{\rm ext}$ of this photon field and temperature $T_{\rm ext}$ of the dusty torus used in the model are poorly constrained but consistent with expectations.  
This model also satisfactorily represents the SED and allows for the choice of parameters very close to equipartition between the magnetic-field and non-thermal electron energy densities.  
The minimum variability time scale is 4.2~hours.  

In addition to the purely leptonic models described above, we also
consider a lepto-hadronic model, in which ultrarelativistic protons
contribute significantly to the high-energy emission through proton-synchrotron 
radiation and $p\gamma$ pion production. The spectra
of $\pi^0$ decay photons as well as the final decay products of
charged pions are evaluated using the templates of \citet{ka08},
accounting for secondary cascades as described
in \citet{boettcher10}. In our model, in addition to the SSC model
outlined above, we assume a power-law distribution of 
relativistic protons, $n(\gamma) \propto \gamma^{-p}$ between
a low- and high-energy cutoff, $\gamma_{\rm p}^{\rm min,max}$,
normalised to a total kinetic luminosity $L_p$ of the proton 
population propagating along the jet. We then evaluate the
energy partition fractions $\epsilon_{Bp} \equiv L_B/L_p$
and $\epsilon_{ep} \equiv L_e/L_p$.  As for the other models, the result is shown
in Figure~\ref{fig_3models}, and the model parameters are listed in
Table~\ref{fitpars}. This model also adequately produces the non-thermal SED. It requires a strongly-magnetically
dominated jet with $\epsilon_{Be} = 1.4 \times 10^6$ and
$\epsilon_{Bp} = 24$.  
Under these conditions, $\gamma$-ray emission is dominated by proton-synchrotron radiation.  
The minimum variability time scale from the size of the emission region is just 4.2~hours.  However, the radiative cooling time of ultrarelativistic protons is on the order of several days, excluding variability on shorter time scales (not yet seen) under this model.

\section{Conclusions}

1ES\,1440+122 was detected by VERITAS at a significance level of 5.5$\sigma$ during the 2008--2010 
observing seasons. In this paper, we described VHE observations of 1ES\,1440+122 
along with the quasi-simultaneous observations with \textit{Swift} in optical, UV, and X-rays 
and \textit{Fermi}-LAT in high-energy $\gamma$-rays. 
The observed non-thermal SED of 1ES\,1440+122 is consistent with purely leptonic
(SSC and EC) models as well as with a hadronic origin.  A leptonic model with an external infrared radiation field as
target for Compton scattering allowed for parameters close to equipartition between the relativistic electron population and the magnetic
field. 
The other models did not allow for partition fractions near unity.  
The model parameters are comparable to those obtained from other studies of VHE BL Lacs.  

Our model SEDs all result in synchrotron peak frequencies contained in the \textit{Swift}-XRT band, very close to $\nu_{\rm{synch}} = 3 \times 10^{17}$~Hz, classifying the source as an HBL according to the scheme of \citet{Nieppola06} or an HSP according to \citet{Abdo:2009iq}.  The Compton-peak frequencies are all close to $\nu_{\rm{Compton}} = 3 \times 10^{25}$~Hz.  The SED is dominated by the lower synchrotron peak (i.e. low Compton dominance), which is generally observed in other VHE HBLs~\citep{Fossati:1998zn, Ghisellini:1998it}.  

Our results show that the SED alone does not allow us to confidently distinguish 
between different models for the high-energy emission from 1ES\,1440+122. Future 
observations, such as probing for intraday variability of VHE $\gamma$-rays, may aid in
distinguishing leptonic and hadronic models. The radiative cooling time scales of
ultrarelativistic protons are of the order of several days, whereas those of 
ultrarelativistic electrons are typically of the order of hours or less. Therefore
rapid (intraday) VHE $\gamma$-ray variability would be an indication of leptonic 
processes dominating the $\gamma$-ray output.  

As a VHE source with a relatively hard spectrum for its redshift, 1ES\,1440+122 could be useful in studies of the EBL.  Several studies have used similar blazar spectra to examine lower limits on the intergalactic magnetic fields \citep{2010Sci...328...73N,2011ApJ...735L..28H,2011ApJ...733L..21D,2012AIPC.1505..606A}.  These studies look for emission in the \textit{Fermi}-LAT band that might have cascaded down from the VERITAS band.  However, there is an ongoing debate in the literature regarding the validity of these limits given the possibility of plasma instability energy losses dominating over the inverse Compton losses, resulting in less lower-energy cascade emission \citep{2012ApJ...752...22B, 2012ApJ...758..101S, 2013ApJ...770...54M}.

\section*{Acknowledgements}
This research is supported by grants from the U.S. Department of Energy Office of Science, the U.S. National Science Foundation and the Smithsonian Institution, by NSERC in Canada, by Science Foundation Ireland (SFI 10/RFP/AST2748) and by STFC in the U.K. We acknowledge the excellent work of the technical support staff at the Fred Lawrence Whipple Observatory and at the collaborating institutions in the construction and operation of the instrument. M. B\"ottcher acknowledges support by the South African Department of Science and Technology through the National Research Foundation under NRF SARChI Chair grant no. 64789.  The VERITAS Collaboration is grateful to Trevor Weekes for his seminal contributions and leadership in the field of VHE gamma-ray astrophysics, which made this study possible.




\bibliographystyle{mnras}
\bibliography{references} 

\begin{thebibliography}{}
\makeatletter
\relax
\def\mn@urlcharsother{\let\do\@makeother \do\$\do\&\do\#\do\^\do\_\do\%\do\~}
\def\mn@doi{\begingroup\mn@urlcharsother \@ifnextchar [ {\mn@doi@}
  {\mn@doi@[]}}
\def\mn@doi@[#1]#2{\def\@tempa{#1}\ifx\@tempa\@empty \href
  {http://dx.doi.org/#2} {doi:#2}\else \href {http://dx.doi.org/#2} {#1}\fi
  \endgroup}
\def\mn@eprint#1#2{\mn@eprint@#1:#2::\@nil}
\def\mn@eprint@arXiv#1{\href {http://arxiv.org/abs/#1} {{\tt arXiv:#1}}}
\def\mn@eprint@dblp#1{\href {http://dblp.uni-trier.de/rec/bibtex/#1.xml}
  {dblp:#1}}
\def\mn@eprint@#1:#2:#3:#4\@nil{\def\@tempa {#1}\def\@tempb {#2}\def\@tempc
  {#3}\ifx \@tempc \@empty \let \@tempc \@tempb \let \@tempb \@tempa \fi \ifx
  \@tempb \@empty \def\@tempb {arXiv}\fi \@ifundefined
  {mn@eprint@\@tempb}{\@tempb:\@tempc}{\expandafter \expandafter \csname
  mn@eprint@\@tempb\endcsname \expandafter{\@tempc}}}

\bibitem[\protect\citeauthoryear{Abdo et~al.}{Abdo et~al.}{2010}]{Abdo:2009iq}
Abdo A.~A.,  et~al., 2010, \mn@doi [Astrophys.J.] {10.1088/0004-637X/716/1/30},
  716, 30

\bibitem[\protect\citeauthoryear{Abdo et~al.}{Abdo et~al.}{2011}]{abdo11}
Abdo A.~A.,  et~al., 2011, \apj, 726, 43

\bibitem[\protect\citeauthoryear{Abramowski et~al.}{Abramowski
  et~al.}{2012}]{Abramowski:2012ux}
Abramowski A.,  et~al., 2012, \mn@doi [A\&A] {10.1051/0004-6361/201117509},
  539, A149

\bibitem[\protect\citeauthoryear{{Abramowski} et~al.,}{{Abramowski}
  et~al.}{2013}]{2013A&A...550A...4H}
{Abramowski} A.,  et~al., 2013, \mn@doi [\aap] {10.1051/0004-6361/201220355},
  \href {http://adsabs.harvard.edu/abs/2013A%26A...550A...4H} {550, A4}

\bibitem[\protect\citeauthoryear{{Abramowski} et~al.}{{Abramowski}
  et~al.}{2014}]{2014A&A...564A...9H}
{Abramowski} A.,  et~al., 2014, \mn@doi [\aap] {10.1051/0004-6361/201322897},
  \href {http://adsabs.harvard.edu/abs/2014A%26A...564A...9H} {564, A9}

\bibitem[\protect\citeauthoryear{Acciari et~al.}{Acciari
  et~al.}{2008}]{Acciari:2008rk}
Acciari V.~A.,  et~al., 2008, \mn@doi [ApJL] {10.1086/592244}, \href
  {http://adsabs.harvard.edu/abs/2008ApJ...684L..73A} {684, L73}

\bibitem[\protect\citeauthoryear{Acciari et~al.}{Acciari
  et~al.}{2009}]{Acciari:2009xz}
Acciari V.~A.,  et~al., 2009, {\apj}, 707, 612

\bibitem[\protect\citeauthoryear{{Acero} et~al.}{{Acero}
  et~al.}{2015}]{2015ApJS..218...23A}
{Acero} F.,  et~al., 2015, \mn@doi [\apjs] {10.1088/0067-0049/218/2/23}, \href
  {http://adsabs.harvard.edu/abs/2015ApJS..218...23A} {218, 23}

\bibitem[\protect\citeauthoryear{{Ackermann} et~al.}{{Ackermann}
  et~al.}{2012a}]{2012ApJS..203....4A}
{Ackermann} M.,  et~al., 2012a, \mn@doi [\apjs] {10.1088/0067-0049/203/1/4},
  \href {http://adsabs.harvard.edu/abs/2012ApJS..203....4A} {203, 4}

\bibitem[\protect\citeauthoryear{{Ackermann} et~al.,}{{Ackermann}
  et~al.}{2012b}]{2012Sci...338.1190A}
{Ackermann} M.,  et~al., 2012b, \mn@doi [Science] {10.1126/science.1227160},
  \href {http://adsabs.harvard.edu/abs/2012Sci...338.1190A} {338, 1190}

\bibitem[\protect\citeauthoryear{{Aharonian} et~al.}{{Aharonian}
  et~al.}{2000}]{2000A&A...353..847A}
{Aharonian} F.,  et~al., 2000, \aap, \href
  {http://adsabs.harvard.edu/abs/2000A%26A...353..847A} {353, 847}

\bibitem[\protect\citeauthoryear{{Aharonian} et~al.}{{Aharonian}
  et~al.}{2006}]{2006A&A...457..899A}
{Aharonian} F.,  et~al., 2006, \mn@doi [\aap] {10.1051/0004-6361:20065351},
  \href {http://adsabs.harvard.edu/abs/2006A%26A...457..899A} {457, 899}

\bibitem[\protect\citeauthoryear{{Arlen} \& {Vassiliev}}{{Arlen} \&
  {Vassiliev}}{2012}]{2012AIPC.1505..606A}
{Arlen} T.~C.,  {Vassiliev} V.~V.,  2012, in {Aharonian} F.~A.,  {Hofmann} W.,
   {Rieger} F.~M.,  eds,  American Institute of Physics Conference Series Vol.
  1505, American Institute of Physics Conference Series. pp 606--609,
  \mn@doi{10.1063/1.4772333}

\bibitem[\protect\citeauthoryear{{Atwood} et~al.}{{Atwood}
  et~al.}{2009}]{2009ApJ...697.1071A}
{Atwood} W.~B.,  et~al., 2009, \mn@doi [\apj] {10.1088/0004-637X/697/2/1071},
  \href {http://adsabs.harvard.edu/abs/2009ApJ...697.1071A} {697, 1071}

\bibitem[\protect\citeauthoryear{{Berge}, {Funk}  \& {Hinton}}{{Berge}
  et~al.}{2007}]{2007A&A...466.1219B}
{Berge} D.,  {Funk} S.,   {Hinton} J.,  2007, \mn@doi [\aap]
  {10.1051/0004-6361:20066674}, \href
  {http://adsabs.harvard.edu/abs/2007A%26A...466.1219B} {466, 1219}

\bibitem[\protect\citeauthoryear{{Biteau} \& {Williams}}{{Biteau} \&
  {Williams}}{2015}]{2015ApJ...812...60B}
{Biteau} J.,  {Williams} D.~A.,  2015, \mn@doi [\apj]
  {10.1088/0004-637X/812/1/60}, \href
  {http://adsabs.harvard.edu/abs/2015ApJ...812...60B} {812, 60}

\bibitem[\protect\citeauthoryear{{Blandford} \& {Levinson}}{{Blandford} \&
  {Levinson}}{1995}]{Blandford95}
{Blandford} R.~D.,  {Levinson} A.,  1995, \apj, 441, 79

\bibitem[\protect\citeauthoryear{{Bloom} \& {Marscher}}{{Bloom} \&
  {Marscher}}{1996}]{Bloom96}
{Bloom} S.~D.,  {Marscher} A.~P.,  1996, \apj, 461, 657

\bibitem[\protect\citeauthoryear{B\"ottcher}{B\"ottcher}{2010}]{boettcher10}
B\"ottcher M.,  2010, in proc. of ``Fermi Meets Jansky'', p.~41

\bibitem[\protect\citeauthoryear{B\"ottcher \& Chiang}{B\"ottcher \&
  Chiang}{2002}]{bc02}
B\"ottcher M.,  Chiang J.,  2002, \apj, 581, 127

\bibitem[\protect\citeauthoryear{{B{\"o}ttcher} et~al.}{{B{\"o}ttcher}
  et~al.}{2013}]{2013ApJ...768...54B}
{B{\"o}ttcher} M.,  et~al., 2013, \mn@doi [\apj] {10.1088/0004-637X/768/1/54},
  \href {http://adsabs.harvard.edu/abs/2013ApJ...768...54B} {768, 54}

\bibitem[\protect\citeauthoryear{{Breeveld} et~al.}{{Breeveld}
  et~al.}{2011}]{2011AIPC.1358..373B}
{Breeveld} A.~A.,  et~al., 2011, in {McEnery} J.~E.,  {Racusin} J.~L.,
  {Gehrels} N.,  eds,  American Institute of Physics Conference Series Vol.
  1358, American Institute of Physics Conference Series. pp 373--376
  (\mn@eprint {arXiv} {1102.4717}), \mn@doi{10.1063/1.3621807}

\bibitem[\protect\citeauthoryear{{Broderick}, {Chang}  \&
  {Pfrommer}}{{Broderick} et~al.}{2012}]{2012ApJ...752...22B}
{Broderick} A.~E.,  {Chang} P.,   {Pfrommer} C.,  2012, \mn@doi [\apj]
  {10.1088/0004-637X/752/1/22}, \href
  {http://adsabs.harvard.edu/abs/2012ApJ...752...22B} {752, 22}

\bibitem[\protect\citeauthoryear{Christiansen et~al.}{Christiansen
  et~al.}{2012}]{HFit12}
Christiansen J.,  et~al., 2012, AIP Conf. Proc., 1505, 709

\bibitem[\protect\citeauthoryear{Costamante \& Ghisellini}{Costamante \&
  Ghisellini}{2002}]{Costamante02}
Costamante L.,  Ghisellini G.,  2002, \aap, 384, 56

\bibitem[\protect\citeauthoryear{Daniel et~al.}{Daniel et~al.}{2007}]{Daniel07}
Daniel M.~K.,  et~al., 2007, 30th ICRC, 3, 1325

\bibitem[\protect\citeauthoryear{{Dermer} et~al.}{{Dermer}
  et~al.}{2011}]{2011ApJ...733L..21D}
{Dermer} C.~D.,  et~al., 2011, \mn@doi [\apjl] {10.1088/2041-8205/733/2/L21},
  \href {http://adsabs.harvard.edu/abs/2011ApJ...733L..21D} {733, L21}

\bibitem[\protect\citeauthoryear{Elvis et~al.}{Elvis et~al.}{1992}]{Elvis92}
Elvis M.,  et~al., 1992, \apjs, 80, 257

\bibitem[\protect\citeauthoryear{{Fitzpatrick}}{{Fitzpatrick}}{1999}]{1999PASP..111...63F}
{Fitzpatrick} E.~L.,  1999, \mn@doi [\pasp] {10.1086/316293}, \href
  {http://adsabs.harvard.edu/abs/1999PASP..111...63F} {111, 63}

\bibitem[\protect\citeauthoryear{Fomin et~al.}{Fomin et~al.}{1994}]{Fomin1994}
Fomin V.~P.,  et~al., 1994, Astropart. Phys., 2, 137

\bibitem[\protect\citeauthoryear{Fossati et~al.}{Fossati
  et~al.}{1998}]{Fossati:1998zn}
Fossati G.,  et~al., 1998, \mn@doi [Mon.Not.Roy.Astron.Soc.]
  {10.1046/j.1365-8711.1998.01828.x}, 299, 433

\bibitem[\protect\citeauthoryear{Franceschini, Rodighiero  \&
  Vaccari}{Franceschini et~al.}{2008}]{Franceschini08}
Franceschini A.,  Rodighiero G.,   Vaccari M.,  2008, \aap, 487, 837

\bibitem[\protect\citeauthoryear{Gehrels et~al.}{Gehrels
  et~al.}{2004}]{Gehrels04}
Gehrels N.,  et~al., 2004, \apj, 611, 1005

\bibitem[\protect\citeauthoryear{Ghisellini et~al.}{Ghisellini
  et~al.}{1998}]{Ghisellini:1998it}
Ghisellini G.,  et~al., 1998, \mn@doi [Mon.Not.Roy.Astron.Soc.]
  {10.1046/j.1365-8711.1998.02032.x}, 301, 451

\bibitem[\protect\citeauthoryear{{Giovannini} et~al.}{{Giovannini}
  et~al.}{2004}]{2004ApJ...613..747G}
{Giovannini} G.,  et~al., 2004, \mn@doi [\apj] {10.1086/423169}, \href
  {http://adsabs.harvard.edu/abs/2004ApJ...613..747G} {613, 747}

\bibitem[\protect\citeauthoryear{Heidt et~al.}{Heidt et~al.}{1999}]{Heidt99}
Heidt J.,  et~al., 1999, \aap, 341, 683

\bibitem[\protect\citeauthoryear{Hillas}{Hillas}{1985}]{Hillas85}
Hillas A.~M.,  1985, 19th ICRC, 3, 445

\bibitem[\protect\citeauthoryear{Holder et~al.}{Holder
  et~al.}{2006}]{Holder:2006gi}
Holder J.,  et~al., 2006, \mn@doi [Astropart.Phys.]
  {10.1016/j.astropartphys.2006.04.002}, 25, 391

\bibitem[\protect\citeauthoryear{Holder et~al.}{Holder et~al.}{2008}]{Holder08}
Holder J.,  et~al., 2008, AIP Conf. Ser., 1085, 657

\bibitem[\protect\citeauthoryear{{Huan} et~al.}{{Huan}
  et~al.}{2011}]{2011ApJ...735L..28H}
{Huan} H.,  et~al., 2011, \mn@doi [\apjl] {10.1088/2041-8205/735/2/L28}, \href
  {http://adsabs.harvard.edu/abs/2011ApJ...735L..28H} {735, L28}

\bibitem[\protect\citeauthoryear{Kalberla et~al.}{Kalberla
  et~al.}{2005}]{Kalberla:2005ts}
Kalberla P.~M.,  et~al., 2005, \mn@doi [A\&A.] {10.1051/0004-6361:20041864},
  440, 775

\bibitem[\protect\citeauthoryear{Kelner \& Aharonian}{Kelner \&
  Aharonian}{2008}]{ka08}
Kelner S.~R.,  Aharonian F.~A.,  2008, \prd, 78, 034013

\bibitem[\protect\citeauthoryear{Li \& Ma}{Li \& Ma}{1983}]{LiMa83}
Li T.~P.,  Ma Y.~Q.,  1983, \apj, 272, 317

\bibitem[\protect\citeauthoryear{Mannheim}{Mannheim}{1998}]{Mannheim98}
Mannheim K.,  1998, Science, 279, 684

\bibitem[\protect\citeauthoryear{{Miniati} \& {Elyiv}}{{Miniati} \&
  {Elyiv}}{2013}]{2013ApJ...770...54M}
{Miniati} F.,  {Elyiv} A.,  2013, \mn@doi [\apj] {10.1088/0004-637X/770/1/54},
  \href {http://adsabs.harvard.edu/abs/2013ApJ...770...54M} {770, 54}

\bibitem[\protect\citeauthoryear{{Neronov} \& {Vovk}}{{Neronov} \&
  {Vovk}}{2010}]{2010Sci...328...73N}
{Neronov} A.,  {Vovk} I.,  2010, \mn@doi [Science] {10.1126/science.1184192},
  \href {http://adsabs.harvard.edu/abs/2010Sci...328...73N} {328, 73}

\bibitem[\protect\citeauthoryear{Nieppola et~al.}{Nieppola
  et~al.}{2006}]{Nieppola06}
Nieppola E.,  et~al., 2006, \aap, 445, 451

\bibitem[\protect\citeauthoryear{Nolan et~al.}{Nolan
  et~al.}{2012}]{Fermi-LAT:2011iqa}
Nolan P.~L.,  et~al., 2012, \apjs, 199, 31

\bibitem[\protect\citeauthoryear{Ong et~al.}{Ong et~al.}{2010}]{Ong10}
Ong R.,  et~al., 2010, A Tel. 2786

\bibitem[\protect\citeauthoryear{Padovani \& Giommi}{Padovani \&
  Giommi}{1995}]{Padovani:1994sh}
Padovani P.,  Giommi P.,  1995, \mn@doi [Astrophys.J.] {10.1086/175631}, 444,
  567

\bibitem[\protect\citeauthoryear{Perkins et~al.}{Perkins
  et~al.}{2009}]{Perkins:2009ia}
Perkins J.,  et~al., 2009, in eConf Proceedings C091122

\bibitem[\protect\citeauthoryear{Pohl \& Schlickeiser}{Pohl \&
  Schlickeiser}{2000}]{Pohl00}
Pohl M.,  Schlickeiser R.,  2000, \aap, 354, 395

\bibitem[\protect\citeauthoryear{{Roming} et~al.}{{Roming}
  et~al.}{2005}]{2005SSRv..120...95R}
{Roming} P.~W.~A.,  et~al., 2005, \mn@doi [\ssr] {10.1007/s11214-005-5095-4},
  \href {http://adsabs.harvard.edu/abs/2005SSRv..120...95R} {120, 95}

\bibitem[\protect\citeauthoryear{Sbarufatti et~al.}{Sbarufatti
  et~al.}{2006}]{Sbarufatti06}
Sbarufatti B.,  et~al., 2006, \aap, 457, 35

\bibitem[\protect\citeauthoryear{Scarpa et~al.}{Scarpa et~al.}{1999}]{Scarpa99}
Scarpa R.,  et~al., 1999, \apj, 521, 134

\bibitem[\protect\citeauthoryear{{Schlafly} \& {Finkbeiner}}{{Schlafly} \&
  {Finkbeiner}}{2011}]{2011ApJ...737..103S}
{Schlafly} E.~F.,  {Finkbeiner} D.~P.,  2011, \mn@doi [\apj]
  {10.1088/0004-637X/737/2/103}, \href
  {http://adsabs.harvard.edu/abs/2011ApJ...737..103S} {737, 103}

\bibitem[\protect\citeauthoryear{{Schlickeiser} et~al.}{{Schlickeiser}
  et~al.}{2012}]{2012ApJ...758..101S}
{Schlickeiser} R.,  et~al., 2012, \mn@doi [\apj] {10.1088/0004-637X/758/2/101},
  \href {http://adsabs.harvard.edu/abs/2012ApJ...758..101S} {758, 101}

\bibitem[\protect\citeauthoryear{Summerlin \& Baring}{Summerlin \&
  Baring}{2012}]{Summerlin:2011xs}
Summerlin E.~J.,  Baring M.~G.,  2012, \mn@doi [Astrophys.J.]
  {10.1088/0004-637X/745/1/63}, 745, 63

\bibitem[\protect\citeauthoryear{Urry \& Padovani}{Urry \&
  Padovani}{1995}]{Urry:1995mg}
Urry C.~M.,  Padovani P.,  1995, \mn@doi [Publ. Astron. Soc. Pac.]
  {10.1086/133630}, 107, 803

\bibitem[\protect\citeauthoryear{de Almeida, Tavecchio  \&
  Mankuzhiyil}{de~Almeida et~al.}{2014}]{deAlmeida:2014iua}
de Almeida U.~B.,  Tavecchio F.,   Mankuzhiyil N.,  2014

\makeatother
\end{thebibliography}







\bsp	
\label{lastpage}
\end{document}